\documentclass[aps,showpacs,pra,twocolumn,superscriptaddress]{revtex4}
\usepackage{epsfig,amssymb,amscd,graphicx}
\usepackage{subfigure}

\begin{document}
\bibliographystyle{apsrev} 
\title{Superballistic Diffusion of Entanglement in Disordered Spin Chains}

\author{J. Fitzsimons}\email{joefitz@thphys.may.ie}
\affiliation{Dept. of Mathematical Physics, National University of Ireland, Maynooth, Co. Kildare, Ireland}
\author{J. Twamley}\email{Jason.Twamley@nuim.ie}
\affiliation{Dept. of Mathematical Physics, National University of Ireland, Maynooth,  Co. Kildare, Ireland}
\affiliation{Centre for Quantum Computer Technology, Macquarie University, Sydney, New South Wales 2109, Australia}

\begin{abstract}
We study the dynamics of a single excitation in an infinite XXZ spin chain, which is launched from the origin. We study the time evolution of the spread of entanglement in the spin chain and obtain an expression for the second order spatial moment of concurrence, about the origin, for both ordered and disordered chains. In this way, we show that a finite central disordered region can lead to sustained superballistic growth in the second order spatial moment of entanglement within the chain.
\end{abstract}

\date{\today}
\maketitle
\subsection{Introduction}
 
Entanglement can be viewed as a physical resource that has no analog in classical information theory. As such, entanglement plays an important role in many quantum information tasks, such as quantum cryptography, teleportation and quantum algorithms. The development of protocols for the distribution of entanglement is an important problem in quantum information processing. Recently several methods have been proposed for accomplishing the related problem of quantum state transfer using spin chains \cite{bose1, subrahmanyam, christandl, albanese, christandl2, osborne, haselgrove, verstraete2, verstraete, plenio, bose2, bose3, korepin8,kay}. An important characteristic in all of these models is that the transfer occurs with minimal control or intervention in the dynamics of the spin chain. Besides achieving perfect quantum state transfer, another important task that can be examined is that of distributing entanglement between a central node and many distributed nodes within a processor. Such a capability may be useful for the generation of multi-qubit entangled states (with the appropriate post-distillation). In any case, gaining a better understanding of how one can use spin chain dynamics to spread out  (rather than transmit) entanglement is of fundamental interest, although this has not seemed to have been addressed much in the literature to date.  

In this letter we consider the second order spatial moment of entanglement in a spin chain, with a single excitation, and show that a finite disordered region can lead to superballistic growth in the the second order spatial moment. The entanglement properties of various spin chains have been examined in a number of recent works \cite{Latorre,jin, korepin1, korepin3, korepin4, korepin5}. It has been shown that, for a particle in a lattice with a finite disordered region, a finite period of superballistic growth can occur in the spatial variance of the wave function \cite{geisel:sswp}. This can be applied to XXZ spinchains, which have nearest neighbor interactions similar to the tight binding approximation used for a 1-dimensional lattice. Unlike the spatial variance, however, the growth in the second order spatial moment of concurrence in a chain with a finite central disordered region is not bounded from above by the similar growth experienced in an ordered chain. The disordered region leads to sustained superballistic spatial growth of the concurrence and provides an efficient mechanism for distributing entanglement along spinchains.

\begin{figure}
\includegraphics[width=3.25in]{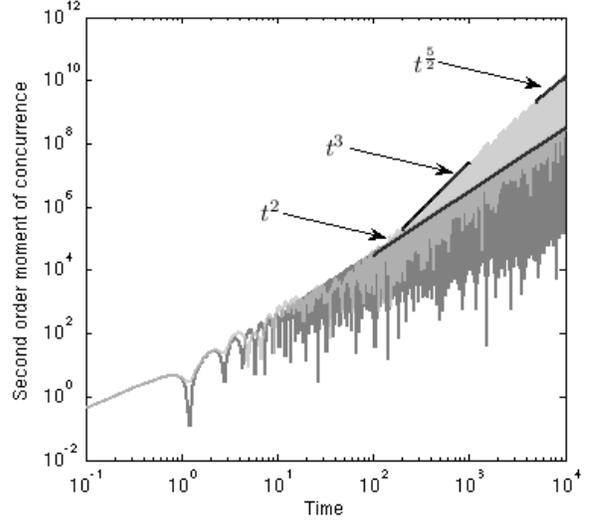}
\caption{The second order spatial moment of concurrence $M(t)$, for an ordered (dark grey) and a disordered (light grey) chain. Each chain is 40500 sites long and the disordered region in the second chain is 100 sites long. The  $J_z$ couplings in the central disordered region  vary randomly, uniformly between 0 and 2.5. A period of super-ballistic, $\sim t^3$, spatial expansion of the moment can be observed. for a period, which tends to a long time super-ballistic expansion rate of $\sim t^{5/2}$, exceeding the long time expansion rate of the ordered chain, $\sim t^2$.}  
\end{figure}

\subsection{Concurrence}

One measure of bipartite entanglement is concurrence \cite{wooters:epqb}, which is a monotonic function of entanglement of formation. For a mixed state, the concurrence between two qubits, $i$ and $j$, is defined to be $C_{ij} = 
\max\{\lambda_1-\lambda_2-\lambda_3-\lambda_4,0\}$, where $\lambda_n$ is 
the square root of the $n^{th}$ eigenvalue of $\rho\tilde{\rho}$ in 
descending order. Here $\tilde{\rho} = (\sigma_y \otimes 
\sigma_y)\rho^\ast(\sigma_y \otimes \sigma_y)$, where $\rho^\ast$ is the 
complex conjugate of $\rho$.

For a single pure excitation of a $N-$spin 1/2 chain, the wave function is 
\begin{equation}
|\psi\rangle = 
\alpha_1| 1 
\rangle + \cdots + \alpha_N |  N \rangle\;\;,\label{wave}
\end{equation}
 where $ |  k\rangle$ is one of $N$ basis states of the $N-$spin chain where the $k^{th}$ site is in the excited state and all other sites are in the ground state. Taking any two spins within this chain $i$, $j$, and tracing over the rest yields the following density matrix:
\[\rho_{ij} = \left( \begin{array}{cccc}
\mu & 0 & 
0 & 0 \\
0 & |\alpha_i|^2 & 
\alpha_i\alpha_j^\ast & 
0 \\
0 & \alpha_j\alpha_i^\ast & 
|\alpha_j|^2 & 
0 \\
0 & 0 & 0 & 0\end{array} \right),\] where $\mu = 1 - |\alpha_i|^2 - |\alpha_j|^2$. 

The eigenvalues of $\rho_{ij}\tilde{\rho}_{ij}$ are 
\mbox{$\{4|\alpha_i|^2|\alpha_j|^2,0,0,0\}$},
so 
\mbox{$\lambda_n 
= \{2 | \alpha_i |  | \alpha_j | ,0,0,0\}$}. Inserting these 
expressions for $\lambda_n$ into 
the previous equation for concurrence gives 
\mbox{$C_{ij}=2 | \alpha_i |  | \alpha_j | $}.

\subsection{Distribution of Entanglement}

One way to study the time evolution of the spatial distribution of entanglement between the origin site and other sites in the spin chain is to look at the second order spatial moment of 
concurrence about the point of origin,
\[M(t)=\sum_{x\not=0} | {x^2}C_{0x}(t) |\;\;, \] where $C_{0x}(t)$ is the 
concurrence at time $t$, between the original site of the excitation, and the site at 
position $x$. $C_{0x}(t)$ can be replaced with the earlier expression for 
the concurrence when only one excitation is acted upon by a spin preserving Hamiltonian.
\begin{eqnarray}
M(t) & = & \sum_{x\not=0} | {x^2}(2 | \alpha_0(t) |  | \alpha_x(t) | ) |\;\;,  \\
  & = & 2 | \alpha_0(t) |  W(t)\;\;,\label{wt}
\end{eqnarray}
where 
\begin{equation}
W(t)\equiv   \sum_{x\not=0}x^2 | \alpha_x(t) |\;\;.
\label{myW}
\end{equation}
 
\subsection{Spin Dynamics: Ordered Chain}
A general Hamiltonian for a spin chain with nearest neighbour interactions and uniform couplings can be written as:
\begin{eqnarray*}
H= \sum_i (J_x^{i,i+1} \sigma_x^i \sigma_x^{i+1} + J_y^{i,i+1}\sigma_y^i\sigma_y^{i+1} + J_z^{i,i+1}\sigma_z^i \sigma_z^{i+1})
\end{eqnarray*}
In such a spin chain, with the additional constraint that couplings obey $J^{i, i+1}_x = J^{i,i+1}_y = \Gamma$, for all sites $i$ in the chain, then the overall $z-$component of spin is conserved. Here we will take $\Gamma=1$, which will give general results up to a rescalling of $t$. For a spin chain of $N$ sites, a basis can be formed from $N$ vectors, $ |  1\rangle... |  N\rangle$, with $ |  k\rangle$ corresponding to a chain with a single excitation at site $k$. Any Hamiltonian, if of the above form in this basis, corresponds to a symmetric tridiagonal matrix. Further, if the coupling parameters $J^i_z = 0$, then the main diagonal will also vanish, giving
\begin{eqnarray*}
H_{ij} & = & \delta (i,j-1) + \delta (i,j+1) \;\;,\\
H & = & L + R\;\;,
\end{eqnarray*}
where $L_{ij} = \delta (i,j-1)$, $R_{ij} = \delta (i,j+1)$. Using this expression for $H$, the evolution operator, $U(t)$ can be found.
\begin{eqnarray*}
U(t) & = & e^{-iHt}
        =  e^{-i(L+R)t}\;\;, \\
       & = & \sum_{n=0}^\infty \frac{1}{n!} (-it)^n (L+R)^n\;\;, \\
       & = & \sum_{n=0}^\infty \frac{1}{n!} (-it)^n \sum_{m=0}^n \frac{n!}{(n-m)!m!}L^m R^{n-m} \;\;,\\
       & = & \sum_{n=0}^\infty (-it)^n \sum_{m=0}^n \frac{1}{(n-m)!m!}D_{n-2m}\;\;,
\end{eqnarray*}
where $D_k = R^k$, and $R^{-1} = L$. The only non-zero entries in $D_k$ are along the $k^{th}$ sub-diagonal, which are all 1. The family of matrices, $\{D_k\}_{-\infty}^{+\infty}$, form a basis for U(t) and one can set,
\begin{eqnarray*}
U(t) & = & \sum_{x=-\infty}^\infty c_x(t) D_x\;\;, \\
c_x(t) & = & \sum_{l=0}^\infty (-it)^{2l+x} \frac{1}{(l+x)!l!}\,, \mbox{ where $2l=n-x$,} \\
& = & (-i)^x t^x \sum_{l=0}^\infty (-1)^l(t)^{2l} \frac{1}{(l+x)!l!} \;\;,\\
& = & (-i)^x J_x(2t) \;\;,\label{myJ}
\end{eqnarray*}
where $J_k$ is the $k^{th}$ order Bessel function of the first kind.

\subsubsection{Upper bound on W(t)}
In the case of the ordered chain we have not been able to obtain exact analytic expressions for (\ref{myW}), but instead we can find bounds for $W(t)$.
Using (\ref{myW}), and the approximation,
\begin{eqnarray}
J_x(z) \approx \left\lbrace\begin{array}{c}
0 \mbox{~for $z > |x|$}\;\;,\\
\sqrt{\frac{2}{\pi z}} \cos(z - \frac{x \pi}{2} - \frac{\pi}{4})\;, \mbox{~for $z \leq |x|$}\\
\end{array}\right.\label{myapprox}
\end{eqnarray}
an upper bound for $W(t)$ can be found,
\begin{eqnarray*}
W(t) & = & \sum_{x=-\infty}^{+\infty} x^2  |  \alpha_x(t)  |  
 =  \sum_{x=-\infty}^{+\infty} x^2  |  J_x(2t)  |  \;\;,\\
& = & \sum_{x=-\infty}^{+\infty}  |  x^2 J_x(2t)  |\;\;. 
\end{eqnarray*}
This can be simplified using the identity $xJ_x(a) = \frac{a}{2}(J_{x-1}(a) + J_{x+1}(a))$ \cite{gradshteyn:tables}, to obtain
\begin{eqnarray*}
W(t) & = & \sum_{x=-\infty}^{+\infty}  |  t^2(J_{x-2}(a) + 2J_x(a) + J_{x+2}(a))\\ 
&   & + t(J_{x-1}(a) - J_{x+1}(a))| \;\;,  \\
& = & (4t^2 + 2t) \sum_{x=-\infty}^{+\infty}  |  J_x(2t)  | \;\;, \\
& \leq & (4t^2 + 2t) \sqrt{\frac{1}{\pi t}}(4t) \;\;,\\
& \approx & \frac{16}{\sqrt{\pi}} t^\frac{5}{2} \mbox{ for $t \gg 1$}\;\;. \\
\\
\end{eqnarray*}

\subsubsection{Lower bound on W(t)} 
A lower bound for $W(t)$ can also be found by observing 
\begin{eqnarray*}  |  J_{2k}(2t)  |  & \geq & J_{2k}(2t) \;\;,\\
\sum_{k = -\infty}^{\infty} (2k)^2  |  J_{2k}(2t)  |  & \geq & \sum_{k = -\infty}^{\infty}(2k)^2 J_{2k}(2t)\;\;, \\
\sum_{x = -\infty}^{\infty} x^2  |  J_{x}(2t)  |  & \geq & \sum_{k = -\infty}^{\infty} (2k)^2  |  J_{2k}(2t)  | \;\;, \\
& \geq & \sum_{k = -\infty}^{\infty}(2k)^2 J_{2k}(2t)\;\;.
\end{eqnarray*}
Using the identities $\sum_{k = -\infty}^{\infty}(2k)^2 J_{2k}(a) = \frac{a^2}{2}$ \cite{gradshteyn:tables}, and (\ref{myW}), this can be rewritten,
\begin{eqnarray*}
W(t) \geq 2 t^2\;\;.\label{myupperbound}
\end{eqnarray*}

Since $2 t^2 \leq W(t) \leq \frac{16}{\sqrt{\pi}} t^\frac{5}{2}$ for $t \gg 1$, then $W(t)$ grows as $t^\lambda$, with $2 \leq \lambda \leq 2.5$.

\subsubsection{Approximation for M(t)}

Using the previous approximation (\ref{myapprox}), which can be rewritten as,
\mbox{$ J_x(z) \approx \sqrt{\frac{2}{\pi z}} (\cos (z - \frac{\pi}{4})\cos (\frac{x \pi}{2}) + \sin (z - \frac{\pi}{4}) \sin (\frac{x \pi}{2}))$,} an expression for $W(t)$ can be found.
\begin{eqnarray*}
W(t) & = & \sum_{x=-\infty}^{\infty} x^2  |  J_x(2t)  |  \, ,\\
& \approx & \sum_{x=-2t}^{2t} x^2 \sqrt{\frac{1}{\pi t}}  |  \cos(2t - \frac{x \pi}{2} - \frac{\pi}{4})  | \,,
\end{eqnarray*}
By taking the average of $|\cos(2t-\frac{x\pi}{2}-\frac{\pi}{4})|$ over a time period from $t=t'$ to $t=t'+\pi$, the time averaged value, $\frac{2}{\pi}$, is obtained. Using this as an approximation for the cosine term, the expression for $W(t)$ reduces to:
\begin{eqnarray*}
W(t)& \approx & \frac{4}{\pi} \sqrt{\frac{1}{\pi t}} \sum_{x=0}^{2t} x^2 \, ,\\
& = & \frac{4}{\pi} \sqrt{\frac{1}{\pi t}} \frac{2t(2t + 1)(4t+1)}{6} \, ,\\
& \to & \frac{32}{3 \pi^\frac{3}{2}} t^\frac{5}{2} \,, \mbox{~as $t \to \infty$}
\end{eqnarray*}

This can be used in the expression for $M(t)$,
\begin{eqnarray*}
M(t) & = & 2  |  \alpha_0(t)  |  W(t) \;\;,\\
& \approx & 2 \sqrt{\frac{1}{\pi t}} \frac{2}{\pi} \frac{32 t^\frac{5}{2} + 24 t^\frac{3}{2} + 4t^\frac{1}{2}}{3 \pi^\frac{3}{2}} \;\;,\\
& \to & \frac{128}{3 \pi^3} t^2\,,\;\; \mbox{~as $t \to \infty$}\;\;.
\end{eqnarray*}

\subsection{Spin Dynamics: Disordered Chain}

It is well known that disorder, which in this case we take to be random $J_z$ coupling strengths, or random magnetic fields in the $z-$direction at each site, leads to spatial localisation of the wave function \cite{anderson}. In a spin chain with a disordered region centred on the initial site of excitation, the wave function will be partially localized, and so the 
amplitude at the initial site will fall off at a lower rate than in the 
case of an ordered chain. We now model the emission of an excitation from the centre site out through a disordered region and into (on either side), a semi-infinite ordered spin chain. 

To consider this we first note that the dynamics of an ordered semi-infinite spin chain, with an  initial single excitation at the first site, can be solved in a similar way to the infinite spin chain.
For an ordered semi-infinite spin chain we have
\begin{equation}
U(t)  
        =  e^{-i(L+R)t} 
        = \sum_{n=0}^\infty \frac{1}{n!} (-it)^n (L+R)^n \;\;.
\end{equation}

For a semi-infinite chain, however, $[L,R] \neq 0$. $(L R)_{i,j} = \delta_{i,j}$, but $(R L)_{i,j} = \delta_{i,j}(1-\delta_{i,1})$. If the initial excitation is at the start of the chain, then the resulting wave, after time $t$, will be the first column of $U(t)$. Taking the first column matrix elements of $U(t)$ to be the wave launched from site $x=0$, as $U(t)_{x\,,0}$, and we can obtain,  
\begin{eqnarray*}
U(t)_{x\,,0} & = & \sum_{n=x}^{\infty} \frac{(-i t)^n}{n!} \left(\left(\begin{array}{c}
n \\
\frac{n-x}{2} \\
\end{array}
\right) - \left(\begin{array}{c}
n \\
\frac{n-x-2}{2} \\
\end{array}
\right)\right) \,,\\
& = & (-i)^x(J_x(2 t) + J_{x+2}(2 t))\;\;, 
\end{eqnarray*}
resulting in a wave function with elements $\alpha_x(t)$ (\ref{wave}), 
\begin{eqnarray*}
\alpha_x(t) & = & (-i)^x(J_x(2 t) + J_{x+2}(2 t)) \,,\\
& = & (-i)^x \frac{(x+1)}{t} J_{x+1}(2 t)\,.
\end{eqnarray*}
We now suppose that the random region, extending from $x=-L$ to $x=+L$, emits an excitation at $+L$,  into the ordered semi-infinite region with amplitude $f(t)$. Shifting coordinates to this interface by setting $x^\prime=x-L$, and then (for convenience) dropping the prime, we can examine the wave propagation into the ordered semi-infinite chain (now $x \geq 1$) to be:
\begin{eqnarray*}
\alpha_{x+L}(t) & = & \int_0^t f(t')(-i)^{x-1} (\frac{x}{t-t'})J_{x}(2(t-t'))dt' \;\;.\\
\end{eqnarray*}
We now break the emission processes into two time bins. For $t \geq t_1$:
\begin{eqnarray}
\alpha_{x+L}(t) & = & \int_0^{t_1} f(t')(-i)^{x-1} (\frac{x}{t-t'})J_{x}(2(t-t'))dt' \nonumber\\
& + & \int_{t_1}^t f(t')(-i)^{x-1} (\frac{x}{t-t'})J_{x}(2(t-t'))dt' \,,\nonumber\\
& = & (-i)^{x-1}(\frac{x}{t-\tau}) J_{x}(2(t-\tau))\int_0^{t_1} f(t') dt' \nonumber\\
& + & \int_{t_1}^t f(t')(-i)^{x-1} (\frac{x}{t-t'})J_{x}(2(t-t'))dt' \,,\nonumber\\
& = & (-i)^{x-1}(\frac{x}{t-\tau}) J_{x}(2(t-\tau))\beta \nonumber\\
& + & \int_{t_1}^t f(t')(-i)^{x-1} (\frac{x}{t-t'})J_{x}(2(t-t'))dt' \,.\nonumber\\ \label{long}
\end{eqnarray}

 For large $t_1$, we will neglect the second term in (\ref{long}), which is equivalent to letting $f(t) = \beta \delta(t-\tau)$. This corresponds to a simple model for the emission in which an excitation is emited, with amplitude $\beta$, at time $t = \tau\ll t_1$. For $t \leq \tau$, there is no emission and $\alpha_{x+L}(t) = 0$, while for $t > \tau$:
\begin{eqnarray}
\alpha_{x+L}(t) & = & \beta(-i)^{x-1} (\frac{x}{t-\tau}) J_{x}(2(t-\tau)) \;\;.\label{diswave}
\end{eqnarray}
The second order moment of concurrence can be split into two parts, $M_o(t)$ and $M_d(t)$. $M_o(t)$ is the contribution to $M(t)$ from the ordered region, and $M_d(t)$ is the contribution from the disordered region. For $t > \tau$:
\begin{eqnarray}
M_o(t) & = & 4 |  \alpha_0(t) | \sum_{x=1}^\infty (x+L)^2\frac{x}{t-\tau} | \beta J_{x}(2(t-\tau)) |  \,,\\
& = & 2 \frac{ | \beta | }{T^\frac{3}{2}\sqrt{\pi}}( |  \cos(2T -\frac{\pi}{4}) |  \sum_{x=1}^T (2x+L)^2 2x \nonumber\\
& & +  |  \sin(2T -\frac{\pi}{4})  |  \sum_{x=1}^T (2x+L-1)^2 (2x-1))\;\;,\nonumber\\
\end{eqnarray}
where $T=t-\tau$.

Since the average value of $ | \sin(z)  |  =  |  \cos(z)  |= \frac{2}{\pi}$, the above can be approximated by
\begin{eqnarray*}
W_o(t) & = & \frac{4  | \beta | }{T^\frac{3}{2}\pi^\frac{3}{2}}\sum_{x=1}^{2T} (x+L)^2 x \,,\\
& = & \frac{2  | \beta | }{T^\frac{3}{2}\pi^\frac{3}{2}}(8T^4 + (\frac{32 L}{3} - 8)T^3 + (4L^2 - 8L + 3)T^2 \\ 
& & + (4L - 2 - 2L^2)T - (18 + \frac{16 L}{3} + 2L^2)) \,,\\
& \approx & \frac{16  | \beta | }{\pi^\frac{3}{2}} (t-\tau)^{\frac{5}{2}}\;, \mbox{~ for $t \gg 0$.}
\end{eqnarray*}

From (\ref{wt}), to obtain $M_o(t)$, we require the amplitude of the wave at the origin or initial site. Since the probability of the excitation entering the semi-infinite chain to the right and left is $ |  \beta  | ^2$, the probability of the excitation remaining within the disordered region is $1-2 |  \beta  | ^2$. If the probability at the initial site is proportional to this, then the amplitude of the wavefunction at the initial origin will be $\gamma \sqrt{1-2 |  \beta  | ^2}$, with $ |  \gamma  |  \leq 1$.
\begin{eqnarray}
M_o(t) & = & \frac{4 | \beta\gamma\sqrt{1-2 | \beta | ^2} | }{T^\frac{3}{2}\pi^\frac{3}{2}}(8T^4 + (\frac{32 L}{3} - 8)T^3 \nonumber\\
& & + (4L^2 - 8L + 3)T^2 + (4L - 2 - 2L^2)T \nonumber\\
& & - (18 + \frac{16 L}{3} + 2L^2)) \,,\nonumber\\
& \approx & \frac{32  | \beta\gamma\sqrt{1-2 | \beta | ^2} | }{\pi^\frac{3}{2}} (t-\tau)^{\frac{5}{2}}\,, \mbox{~ for $t \gg 0$.}\nonumber\\
\end{eqnarray}

There is only a finite disordered range, so $W_{d}(t)$ is bounded
from above by $2L^2$. This in turn leads to an upper limit on
$M_{d}(t)$ of $4  | \gamma\sqrt{1-2 | \beta | ^2} |  L^2$. Since $M_{o} \propto t^\frac{5}{2}$ and $M_{d} \leq 2 \beta L^2$, 
\begin{equation}
M(t) \approx M_{o}(t) \approx 
\frac{32  | \beta\gamma\sqrt{1-2 | \beta | ^2} | }{\pi^\frac{3}{2}} (t-\tau)^{\frac{5}{2}}, \mbox{ for $t \gg 0$}\;\;.
\end{equation}

In summary, we have examined the emission of a single excitation from a central site on a spin chain in the two cases where the chain is ordered and where the origin site is surrounded by a finite region of disorder. We have found that the spatial expansion of the two-site entanglement, between the origin site  
and other sites on the chain, is much more rapid in the latter case where disorder is present, than in the ordered case. Indeed the second order moment of the spatial extent of the concurrence expands super-ballistically. This effect could be used to rapidly distribute entanglement throughout regions of a spin chain. If, rather than distribute entanglement throughout a chain, one wished to spatially separate an entangled state one might instead appeal to a recent scheme outlined in \cite{kay}.

J. Fitzsimons acknowledges support from the Embark Initiative while J. Twamley acknowledges support from Science Foundation Ireland.

\bibliography{Superballistic}
\end{document}